\documentstyle[twoside,fleqn,npb,epsfig]{article}
\def\beq{\begin{equation}}
\def\eeq{\end{equation}}
\def\beqa{\begin{eqnarray}}
\def\eeqa{\end{eqnarray}}

\newcommand{\AmS}{{\protect\the\textfont2
  A\kern-.1667em\lower.5ex\hbox{M}\kern-.125emS}}

\title{NNLO results for heavy quark and direct photon production 
near threshold}

\author{Nikolaos Kidonakis \address{Department of Physics, 
        Florida State University, \\
        Tallahassee, FL 32306-4350, USA}
\thanks{This work was supported in part by the U.S. Department of Energy.}}

\begin{document}

\begin{abstract}
I present some results for the next-to-next-to-leading order expansions of the 
resummed cross sections for heavy quark and direct photon production 
near threshold in hadronic collisions.
\end{abstract}

% typeset front matter (including abstract)
\maketitle

\section{INTRODUCTION}

Heavy quark and direct photon production in hadronic
collisions are processes of great interest. The discovery of the top
has increased efforts in the calculation of heavy quark production 
cross sections while direct photon production is important for
determinations of gluon distributions. Near threshold for the production
of the final state in these processes one can resum logarithmic 
corrections originating from soft gluon emission.
A formalism and methods for the 
resummation of soft-gluon contributions at next-to-leading logarithmic
(NLL) accuracy for a variety of QCD hard scattering processes, 
including heavy quark, jet, and direct photon production,
have been recently developed (for a review see Ref. \cite {NK}). 
The resummed cross sections have been expanded at next-to-next-to-leading 
order (NNLO) and exhibit a substantially reduced scale dependence relative 
to next-to-leading order (NLO).  

\section{NNLO RESULTS FOR HEAVY QUARK PRODUCTION}

The heavy quark production cross section in hadronic collisions
may be written as a convolution of parton distribution functions
$\phi$ with the partonic hard scattering $\hat{\sigma}$:
\beq
\sigma_{h_1h_2\rightarrow Q{\bar Q}}
=\sum_f
\phi_{f/h_1} \otimes \phi_{{\bar f}/h_2}
\otimes\hat{\sigma}_{f{\bar f}\rightarrow Q{\bar Q}}\, .
\label{sigmaconv}
\eeq
We now define variables $z \equiv Q^2/s$ and $\tau \equiv z_{\rm min}=Q^2/S$,
where $Q^2$, $s$ and $S$ are 
the invariant masses squared of the heavy quark pair, and the incoming 
partons and hadrons, respectively. Then, near threshold, 
$z=1$, $\hat{\sigma}$ includes 
logarithmic terms of the form $-({\alpha_s}^n/n!)[(\ln^m(1-z))/(1-z)]_{+}$, 
with $m \le 2n-1$ at $n$th order in $\alpha_s$, which can be resummed to all 
orders in perturbative QCD.
 
If we take moments of Eq. (\ref{sigmaconv}), the convolution
becomes a simple product.
Replacing the incoming hadrons by partons we can then write moments
of the partonic cross section as 
\beqa
{\tilde{\sigma}}_{f{\bar f}\rightarrow Q{\bar Q}}(N)
&=&{\tilde{\phi}}_{f/f}(N)\,  {\tilde{\phi}}_{{\bar f}/{\bar f}}(N)\,
\hat{\sigma}_{f{\bar f}\rightarrow Q{\bar Q}}(N)
\nonumber \\ && \hspace{-13mm}
={\tilde{\psi}}_{f/f}(N)\, {\tilde{\psi}}_{{\bar f}/{\bar f}}(N)\, 
H_{IJ} \, {\tilde{S}}_{JI}(N) \, ,
\label{sigmamom}
\eeqa
where the moments are defined by \cite{KS} 
$\tilde{\sigma}(N)=\int_0^1 d\tau \, 
\tau^{N-1}\sigma(\tau)$, ${\hat\sigma}(N)=\int_0^1 dz \, 
z^{N-1}{\hat\sigma}(z)$, and
$\tilde{\phi}(N)=\int_0^1dx\, x^{N-1}\phi(x)$.
Under moments the plus distributions in $1-z$ produce powers of $\ln N$.
In the second line of Eq. (\ref{sigmamom}) we have introduced a
refactorization \cite{KS} in terms of center-of-mass parton 
distributions $\psi$, a soft gluon function $S$ which describes 
noncollinear soft gluons, and the $N$-independent hard scattering $H$. 
The indices $I,J$ describe color exchange
and the functions $H$ and $S$ are matrices in color space.

Using Eq. (\ref{sigmamom}), we then have
\beq
\hat{\sigma}_{f{\bar f}\rightarrow Q{\bar Q}}(N)
=\left[\frac{{\tilde{\psi}}_{f/f}(N)}
{{\tilde{\phi}}_{f/f}(N)}\right]^2 H_{IJ} \; {\tilde{S}}_{JI}(N) \, .
\eeq

Resummation follows from the renormalization properties of these 
functions as described in Refs. \cite{NK,KS}.
The resummed heavy quark cross section in moment space is then
\beqa
{\hat{\sigma}}_{f{\bar f}\rightarrow Q{\bar Q}}(N) &=&
\exp \left\{ 2 \left[ E^{(f)}(N)+E^{(f)}_{\rm scale}\right] \right\}
\nonumber \\ && \hspace{-28mm} \times \; 
H_{IJ}\left({Q\over\mu},\alpha_s(\mu^2)\right) \;
{\tilde S}_{JI} \left(1,\alpha_s(Q^2/N^2) \right) 
\nonumber \\ && \hspace{-28mm} \times \,
\exp\left[\int^{Q/N}_{\mu}\frac{d \bar{\mu}}{\bar{\mu}}
[\lambda_I(\alpha_s(\bar{\mu}^2))
+\lambda^*_J(\alpha_s(\bar{\mu}^2))]\right] \, . 
\nonumber \\ 
\label{resHQ}
\eeqa
The function $E^{(f)}$ resums the $N$-dependence of the ratio 
$\psi_{f/f}/\phi_{f/f}$ \cite{GS,KS} while the scale variation is given by 
$E^{(f)}_{\rm scale}$.
In the last exponent, the $\lambda's$ are eigenvalues of the soft 
anomalous dimension matrix $\Gamma_S$, which is determined from 
renormalization group analysis of the soft function $S$ and has been 
calculated explicitly at one-loop in Ref. \cite{KS}.  
For finite-order expansions one does not need to diagonalize $\Gamma_S$;
this is a simplification compared to the full resummed cross 
section \cite{NKJSRV}.
Explicit results for all the exponents in the above expression are
given in Refs. \cite{NK,KS}. 

The NLO expansions for both the $q {\bar q} \rightarrow Q {\bar Q}$
and $gg \rightarrow Q {\bar Q}$ partonic channels have been obtained in 
Refs. \cite{NK,KS}.
These results are in agreement with the one-loop results
in Ref.~\cite{mengetal}.

The expansion of the resummed cross section at NNLO provides us
with the dominant terms near threshold at that order.
For the $q {\bar q} \rightarrow Q {\bar Q}$ channel in the 
$\overline {\rm MS}$ scheme, we find the following NNLO corrections 
in pair inclusive kinematics:
\begin{eqnarray}
{\hat \sigma}^{\overline {\rm MS} \, (2)}_{q{\bar q}\rightarrow Q{\bar Q}}
(1-z,m^2,s,t_1,u_1)&=&
\sigma^B_{q{\bar q}\rightarrow Q{\bar Q}}
\frac{\alpha_s^2}{\pi^2}
\nonumber \\ && \hspace{-52mm} \times 
\left\{8 C_F^2 \left[\frac{\ln^3(1-z)}{1-z}\right]_{+} \! 
+\left[\frac{\ln^2(1-z)}{1-z}\right]_{+} \! C_F \left[-\beta_0\right. \right.
\nonumber \\ && \hspace{-51mm}
{}+12C_F \left(4 \ln\left(\frac{u_1}{t_1}\right)-{\rm Re}\,
L_{\beta}-1-\ln\left(\frac{\mu^2}{s}\right)\right) 
\nonumber \\ && \hspace{-52mm} \left. \left.
{}+6 C_A \left(-3 \ln\left(\frac{u_1}{t_1}\right)
-\ln\left(\frac{m^2s}{t_1 u_1}\right)
+{\rm Re} \, L_{\beta}\right)\right]\right\}
\nonumber \\ && \hspace{-45mm}
{}+ \cdots
\end{eqnarray}
where $\sigma^B_{q{\bar q}\rightarrow Q{\bar Q}}$ is the Born cross section,
$m$ is the heavy quark mass, $t_1=(p_q-p_{\bar Q})^2-m^2$, 
$u_1=(p_{\bar q}-p_{\bar Q})^2-m^2$,
$\beta_0=11-2n_f/3$, with $n_f$ the number of flavors, 
and we have omitted subleading powers of $\ln(1-z)$.

Analogous results have been obtained in the DIS scheme, and also
in single-particle inclusive kinematics in both schemes~\cite{KLMV}.
The NNLO corrections for the $gg \rightarrow Q {\bar Q}$ channel
are lengthier and are presented in Refs. \cite{NK,KLMV}.
These results are useful not only for the total cross section but also
for the calculation of heavy quark differential distributions \cite{pty}.

\section{NNLO RESULTS FOR DIRECT PHOTON PRODUCTION}

Next, we discuss direct photon production in hadronic collisions.
The factorized cross section may again be written as a convolution
of parton distributions with the hard scattering
\beq
\sigma_{h_1h_2\rightarrow\gamma}=
\sum_f \phi_{f_1/h_1} \otimes  
\phi_{f_2/h_2} \otimes {\hat \sigma}_{f_1f_2 \rightarrow \gamma} \, .
\eeq
If we take moments of the above equation
and replace the incoming hadrons by partons, we can write moments
of the partonic cross section as 
\beqa
{\tilde{\sigma}}_{f_1f_2\rightarrow \gamma}(N) \! \!
&=& \! \! {\tilde{\phi}}_{f_1/f_1}(N)\,  {\tilde{\phi}}_{f_2/f_2}(N)\,
\hat{\sigma}_{f_1 f_2 \rightarrow \gamma}(N)
\nonumber \\ && \hspace{-17mm}
={\tilde{\psi}}_{f_1/f_1}(N)\, {\tilde{\psi}}_{f_2/f_2}(N)\, 
{\tilde{J}}(N) \, H \, {\tilde{S}}(N) \, ,
\label{sigmamomdp}
\eeqa
where in the second line of Eq. (\ref{sigmamomdp}) 
we have introduced a refactorization
as described for heavy quark production. Here, we have in addition 
a function $J$ which describes the outgoing jet \cite{KOS,LOS}.

After solving for $\hat{\sigma}_{f_1 f_2\rightarrow \gamma}(N)$ 
in Eq. (\ref{sigmamomdp}) and resumming the $N$-dependence of all 
functions, we may write the resummed cross section as
\beqa
{\hat{\sigma}}_{f_1 f_2 \rightarrow \gamma}(N) \! \! \! &=& \! \! \!
\exp \left \{ \sum_{i=1,2} \left [E^{(f_i)}(N_i)
+E^{(f_i)}_{\rm scale}\right]\right\}
\nonumber \\ && \hspace{-26mm} \times \;
\exp \left \{E'_{(f_J)}(N) \right\} \,
H\left(\alpha_s(\mu^2)\right) \, S\left(1, \alpha_s(S/N^2)\right) 
\nonumber \\ && \hspace{-26mm} \times \; \left.
\exp \left[\int_\mu^{\sqrt{S}/N} {d\mu' \over \mu'} \,
2 \, {\rm Re} \Gamma_S\left(\alpha_s(\mu'^2)\right)\right]\right\}.
\eeqa
The exponent $E'_{(f_J)}$ resums the $N$-dependence of the outgoing
jet \cite{KOS,LOS} while $E^{(f_i)}$ resums the ratio $\psi/\phi$
as we discussed in the previous section. We note that because of the 
simpler color structure of the hard scattering, here $H$, $S$,
and $\Gamma_S$ are simple functions; thus we drop the color indices.   

The threshold region is given in terms of the variable $w \equiv -u/(s+t)$
by $w=1$. We also define $v \equiv 1+t/s$ with $s=(p_1+p_2)^2$, 
$t=(p_1-p_{\gamma})^2$, and $u=(p_2-p_{\gamma})^2$.
The NLO $\overline {\rm MS}$ expansion of the resummed cross section
is presented in Refs. \cite{LOS,NKJO}. Agreement is found with the exact NLO
cross section in Ref. \cite{GV}.

\begin{figure}
\centerline{
\psfig{file=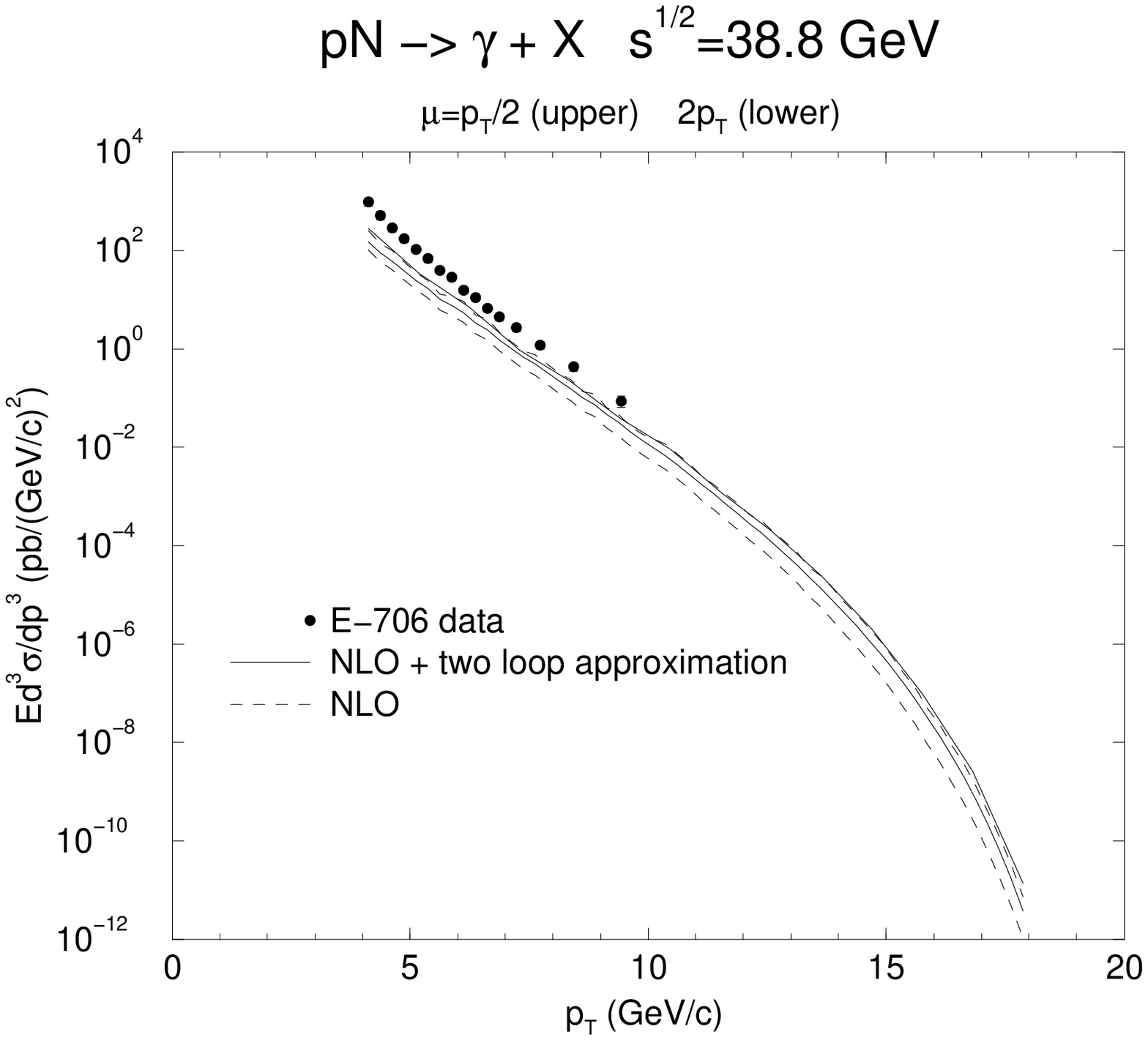,height=2.9in,width=2.9in,clip=}}
{Figure 1. NLO and NNLO results for direct photon production in hadronic 
collisions.}
\label{fig1}
\end{figure}

The NNLO $\overline {\rm MS}$ corrections for the partonic subprocess 
$q(p_1)+ g(p_2) \rightarrow q(p_J)+\gamma(p_{\gamma})$ are  
\beqa
{\hat \sigma}^{\overline {\rm MS}\, (2)}_{qg \rightarrow q\gamma}
(1-w,s,v)
&=&\sigma^B_{qg\rightarrow q\gamma} \frac{\alpha_s^2}{\pi^2} 
\nonumber \\ && \hspace{-38mm} \times \,
\left\{\left(\frac{C_F^2}{2}+2C_F C_A+2C_A^2\right)
\left[\frac{\ln^3(1-w)}{1-w}\right]_{+} \right.
\nonumber \\ && \hspace{-36mm}
{}+\left[\frac{3}{2} C_F^2\left(-\frac{3}{4}+\ln v
-\ln\left(\frac{\mu^2}{s}\right)\right) \right.
\nonumber \\ && \hspace{-34mm}
{}+3C_A^2\left(\ln\left(\frac{v}{1-v}\right)
-\ln\left(\frac{\mu^2}{s}\right)\right)
\nonumber \\ && \hspace{-34mm} 
{}+\frac{3}{2}C_F C_A\left(-\ln(1-v)+3\ln v 
-3\ln\left(\frac{\mu^2}{s}\right)\right.
\nonumber \\ && \hspace{-34mm} \left. \left. \left.
{}-\frac{3}{2}\right)-\frac{\beta_0}{2}\left(\frac{C_F}{4}+C_A\right)\right]
\left[\frac{\ln^2(1-w)}{1-w}\right]_{+} \right\} 
\nonumber \\ && \hspace{-30mm}
{} + \cdots 
\eeqa
where $\sigma^B_{qg\rightarrow q\gamma}$ is the Born cross section for this
channel and we have omitted subleading powers of $\ln(1-w)$.
Analogous results have been obtained for the partonic channel
$q(p_1)+ {\bar q}(p_2)  \rightarrow g(p_J)+\gamma(p_{\gamma})$
\cite{NK,NKJO}.

In Fig. 1 we show some numerical results \cite{NKJO}
for direct photon production
and compare with the experimental results from the E706 Collaboration
at Fermilab \cite{E706}.
We see that the sum of the exact NLO cross section and NNLO
approximate corrections shows a much reduced
dependence on the factorization scale relative to the exact NLO
cross section alone. However, the NNLO cross section is still  
below the E706 data.

\end{document}